\documentclass[sigconf,nonacm,natbib=true]{acmart}

\usepackage{tikz}
\usetikzlibrary{arrows.meta, positioning, shadows.blur}
\usepackage{longtable}
\usepackage{booktabs}
\usepackage{array}
\usepackage{calc}
\usepackage{placeins}

\setcopyright{none}
\renewcommand\footnotetextcopyrightpermission[1]{}
\title{Human-in-the-Loop Nugget Annotation for \\ Accountable LLM-as-a-Judge Evaluations}

\author{Laura Dietz}
\affiliation{%
  \institution{University of New Hampshire}
  \country{USA}
}
\email{dietz@cs.unh.edu}

\begin{abstract}

Evaluating AI/Agentic system outputs reliably requires human judgment, but how one
incorporates the human determines whether one gets a real quality signal
or expensive theater. The common approaches either accidentally anchor human experts (leading to rubber-stamping) or leave them
unsupported in high-variance labeling tasks. We present a prototype
annotation tool that implements a different division of labor: humans
identify what information matters (nuggets), while LLMs handle
high-volume matching of nuggets to system outputs. This plays to each
party's strengths while maintaining genuine human oversight. We describe
the three-phase workflow, key
design decisions, and how exported nugget banks integrate with automated
judges.\footnote{Web demo: \url{https://trec-auto-judge.cs.unh.edu/annotate/nugget-hil-demo.html}}

\end{abstract}


\begin{document}

\maketitle

\section{Introduction}\label{the-problem}

Using an LLM to judge another LLM's output leads to circular effects that render the evaluation invalid. When the judge
shares biases, training data, or architectural patterns with the system
being evaluated, the resulting scores conflate ``sounds like what an LLM
would write'' with ``actually addresses the user's information need well.'' 

This failure mode is not related to the quality of the underlying LLM. It is merely a result from using the same approach to obtain results and to evaluate---akin to a fifth-grader grading their own essays. The effect has been empirically confirmed \cite{clarke2024llm,dietz2026insider}.

% \paragraph{How to not Incorporate the Human}\label{two-common-fixes-that-fail}
Two common ways to respond to this circularity both fail in different ways:

\paragraph{Approach 1: Human Verification of AI Proposals.} The LLM proposes a decision, then a human reviews and corrects when disagreeing. This
triggers anchoring bias \cite{tversky1974judgment}. The problem is that the human sees the
machine's answer before forming their own opinion. Research shows this
pattern leads to blind agreement even when the judge is wrong \cite{agudo2024impact,fok2024search}. This problem is described as the Rubber-Stamp Effect (Judge Trope \#12, \cite{dietz2025principles}): humans under time pressure tend to believe the AI rather than
providing critical oversight. When such contaminated judgments are used to train or calibrate an LLM judge, the errors compound due to concept drift.

\paragraph{Approach 2: Manual Test Sets.} Humans provide relevance assessments on examples
without seeing AI scores, then those assessments replace, calibrate, or evaluate
the judge. This approach is a safe way to remove anchoring but leaves the human expert completely unsupported. Assigning a single
numerical quality score to a lengthy response for a complex task is a cognitively demanding, 
high-effort task. This is referred to as Black-box Labeling (Judge Trope \#13 \cite{dietz2025tropes}): when the criterion is
complex, the label becomes hard to produce and hard to interpret.
\citet{shankar2024validates} document criteria drift, where
annotators refine their standards as they work, making the benchmark
unreliable.
It is well-known that even highly trained human experts suffer from fatigue, provide inconsistent assessments, and are missing important details. As a result, fully manual benchmarks tend to contain many data annotation errors that negatively affect the evaluation results.

\paragraph{Approach 3: Humans Specify Criteria, AI Applies Them at Scale.}
In this paper we adopt a third approach: human experts decide what information matters, and the AI is restricted to finding whether those human-authored information needs are expressed in system outputs. This preserves the central accountability of human judgment while using the LLM for the narrower linguistic task it performs well: matching semantically equivalent statements across many responses. The key design principle is therefore not to ask the LLM to decide what is relevant, but to let it operationalize criteria that humans have already articulated.

\begin{figure*}
\centering
\includegraphics[scale=0.35,keepaspectratio]{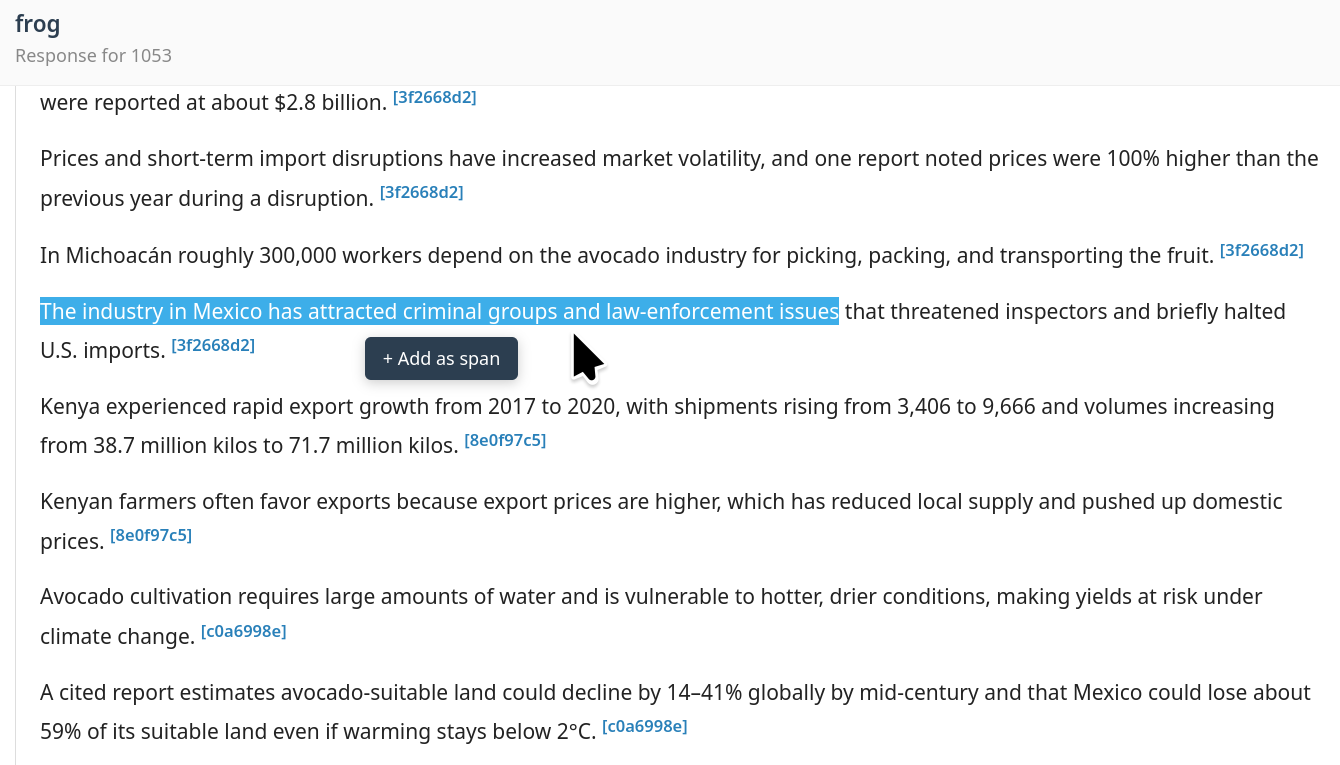}
\includegraphics[scale=0.35,keepaspectratio]{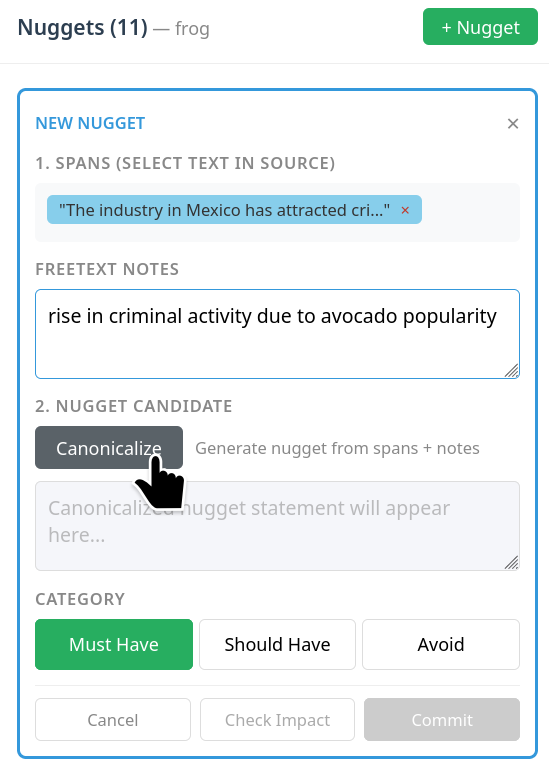}
\caption{Grounding and note-taking for manual nugget curation. The human expert selects ``The industry in Mexico has attracted criminal groups and law-enforcement issues'' and types the free-text note: ``rise in criminal activity due to avocado popularity.'' These notes describe their intent of an important piece of information before any AI involvement.
\label{fig:nugget-grounding}}
\end{figure*}

\section{Nugget-based LLM Judges}

Our approach builds on prior work on Nugget-based LLM Judges. A nugget judge evaluates a system response by checking whether it contains a set of predefined atomic information needs, rather than asking for one holistic relevance or quality score. These pieces of information are commonly referred to as ``nuggets'' or ``SCUs'', and describe facts, constraints, or errors that an ideal response should contain or avoid. Unlike holistic relevance scores, nuggets are:

\begin{itemize}
% \tightlist
\item
  \textbf{Explicit.} The evaluator articulates what matters, not just
  whether the response feels good.
\item
  \textbf{Verifiable.} Each nugget can be checked independently against
  a response.
\item
  \textbf{Reusable.} The same nugget bank evaluates all systems on a
  query and can be repeated.
\item
  \textbf{Auditable.} Disagreements can be traced to specific
  information pieces.
\end{itemize}

A collection of nuggets for a query forms a nugget bank. Each nugget in the bank is assigned an importance category that determines how it contributes to the final evaluation:

\begin{description}
    \item[Must Have:] Critical information. Responses that lack this information are not acceptable.
    \item[Should Have:] Important and nice-to-have information.
    \item[Avoid:] Information that is wrong, off-topic, or otherwise undesirable---an anti-nugget.
\end{description}

The role of the LLM judge is then restricted to matching: for every system response, the judge determines whether the response expresses each nugget, and records the matching grade and supporting evidence. Must-have and should-have nuggets reward responses that cover the desired information, while avoid nuggets penalize responses that include undesirable information. This turns a complex evaluation task into many smaller, auditable matching decisions.

Once all system responses are graded against all nuggets, the nugget bank can be used to compute commonly used nugget-based evaluation metrics. Average nugget grade summarizes how strongly a response satisfies the applicable nuggets. Nugget coverage measures how many must-have and should-have nuggets are addressed. Weighted scores combine nugget grades with the must/should/avoid importance categories, allowing critical omissions or harmful inclusions to matter more than missing optional details. These metrics are established ways to compare system quality in nugget-based evaluation \cite{farzi2024pencils,walden2025auto}.

The advantage of this process is that human experts are intellectually in charge of making decisions about what must be included in an ideal system response. At the same time, AI is providing scale and consistency---and even repeatability. Hence, both humans and AI are playing to their respective strengths. 

\section{Our Approach: Supporting Human Opinion Formation}\label{our-approach}

Given this division of labor, the remaining design question is how to support humans while they identify nuggets. In our design, accountability means that every evaluation criterion can be traced to a human action: a selected span, a free-text note, a category assignment, or an explicit edit. The interface should help experts form their own judgments first, then use AI only for narrowly scoped assistance where humans are prone to inconsistency, omission, or fatigue. The LLM may formalize or preview the consequences of a human-authored nugget, but it cannot introduce evaluation criteria on its own.

\subsection{Supporting Accountable Nugget Curation}

Humans recognize a good response when they see it, but find it difficult to give an all-encompassing list of must-haves without grounding. To help the human in recognizing relevant information, concrete system output is displayed. This will feel like ``eyeballing'' AI output, but we ask the human expert to perform a light-weight form of highlighting relevant text spans or taking free-form notes. These are used as input for nugget formation as shown in Figure \ref{fig:nugget-grounding}.

Since this grounding can also be a potential source of bias, it is important to have the expert inspect system responses in a random order and to anonymize the system names. This avoids that humans prefer content in a system that is a favored candidate for deployment. The order consideration should be tracked and carefully considered in a quality control step.

% nugget grounding picture

We suggest to have the human expert choose whether highlighting text spans and/or free-text is more appropriate in aiding the nugget curation process. During curation, the expert also assigns each nugget to the must-have, should-have, or avoid category introduced above.

\subsection{Nugget Canonicalization Support}

Finally, when teams of human experts work together on creating nuggets, there tends to be variation due to different styles of writing. For AI to offer matching support, it is best if all nuggets are somewhat similarly phrased and if the nugget formulation works well with the matching prompt.

% canonicalization picture
\begin{figure}
\centering
\includegraphics[width=1\columnwidth,keepaspectratio,alt={Step 1-4}]{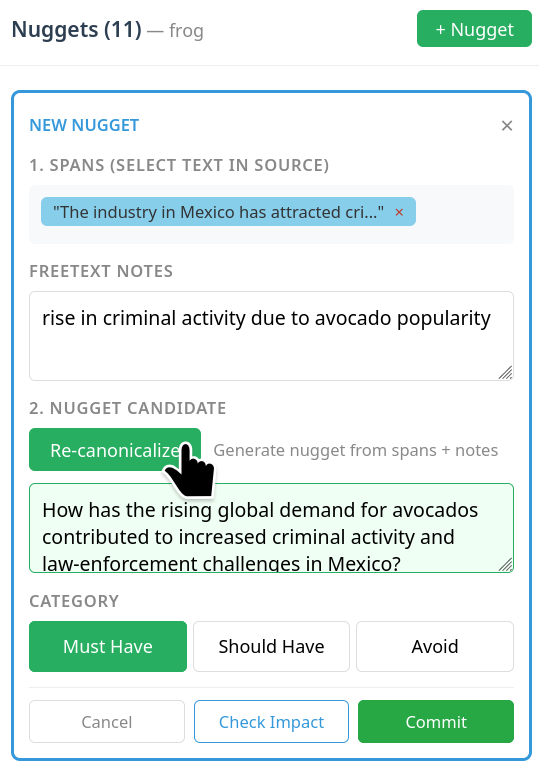}
\caption{Nugget canonicalization support. After clicking ``Canonicalize'', the LLM converts selected spans, free-text, and the task description into one nugget. In this prototype, the alignment prompt expects an open-ended question such as: ``How
has the rising global demand for avocados contributed to increased
criminal activity and law-enforcement challenges in Mexico?''. The human expert can further edit the proposed phrasing. 
\label{fig:nugget-canonicalization}
}
\end{figure}

Based on human artifacts, such as highlighted text spans and free-text notes, the AI can assist the canonical phrasing of the nuggets, and ensure that nuggets are atomic and self-contained and make unambiguous reference to the query. This process is depicted in Figure \ref{fig:nugget-canonicalization}. 

It is important that the AI's role is restricted to formalization. The human expert decides what is essential information; the LLM merely helps express it as a verifiable question. Hence the AI is not permitted to propose what is relevant.

While it is possible for an LLM to propose nuggets without human input \cite{farzi2024pencils,pradeep2024autonuggetizer,li2026dogmatiq,dietz2026too}, it significantly weakens the guardrail against circularity \cite{dietz2026insider} and limits whether the human experts are genuinely accountable for the resulting decisions.

\subsection{Nugget Impact Feedback}

How a particular nugget is phrased has consequences of how well the AI is able to find all the relevant matches in system responses. If the formulation is too specific, no matches will be found, but when it is too generic, the nugget will not distinguish quality differences among the best systems.
To help the human expert choose the appropriate formulation, we include a ``Check Impact'' feature: As soon as the expert is phrasing a nugget, the AI will identify quotes of system output that would match the nugget phrasing according to the nugget-matching prompt. 

% impact image

\begin{figure}
\centering
\includegraphics[width=1\columnwidth,keepaspectratio,alt={Step 2-2}]{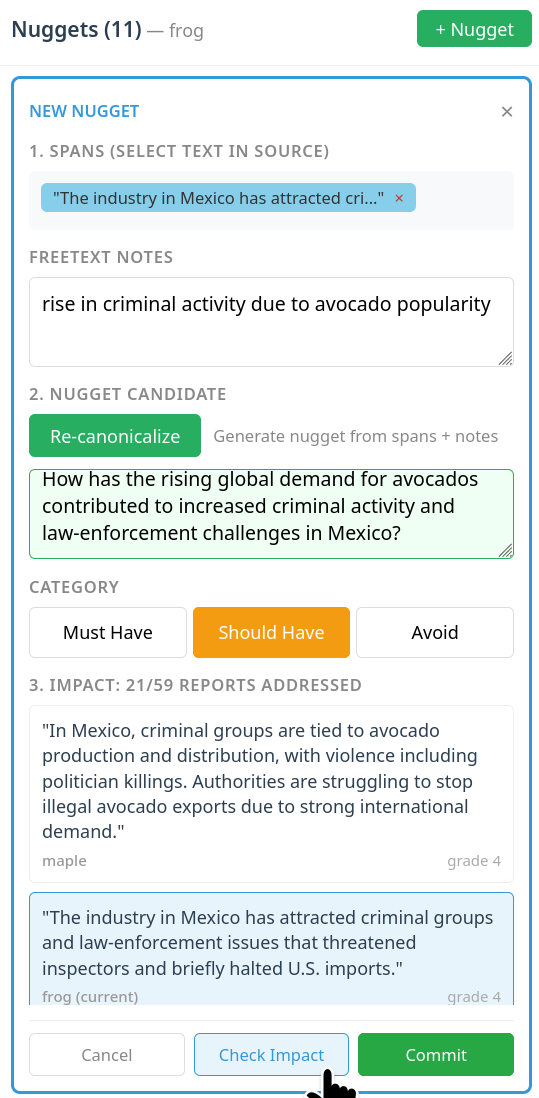}
\caption{Check Nugget Impact. After clicking Check Impact, the interface shows
in how many system outputs this nugget was located, along with supporting quotes. Here the system ``maple''
(grade 4) shows: ``In Mexico, criminal groups are tied to avocado
production and distribution, with violence including politician
killings\ldots{}''. The annotator sees exactly how this nugget would affect system quality measurements. Here the user receives confirmation that the system response that inspired the nugget is also captured before committing the nugget. \label{fig:check-impact}}
\end{figure}

As demonstrated in Figure \ref{fig:check-impact}, quotes of matching system responses are displayed along with the nugget coverage grade (5 for perfect coverage, 1 for only topical references). A click on the quote will display the quote in context of the system response. For nuggets that are grounded with highlighted spans in a particular system output, the user will receive feedback on whether the highlighted text would be matched by the nugget alignment prompt.

Based on inspection of the results and match statistics, the human expert is invited to change the formulation of the nugget to obtain the intended coverage breadth and specificity.

\subsection{Quality Control}

After nugget matching, the quality-control view lets the human expert inspect whether the resulting leaderboard agrees with their informed judgment of the systems. Rather than treating the scores as final, the interface exposes the effect of the current nugget bank: experts can adjust category weights, select subsets of related nuggets, use solo-mode to isolate individual nuggets, and check whether avoid nuggets are penalizing responses as intended.

\begin{figure}
\centering
\includegraphics[width=1\columnwidth,keepaspectratio,alt={Step 3-4}]{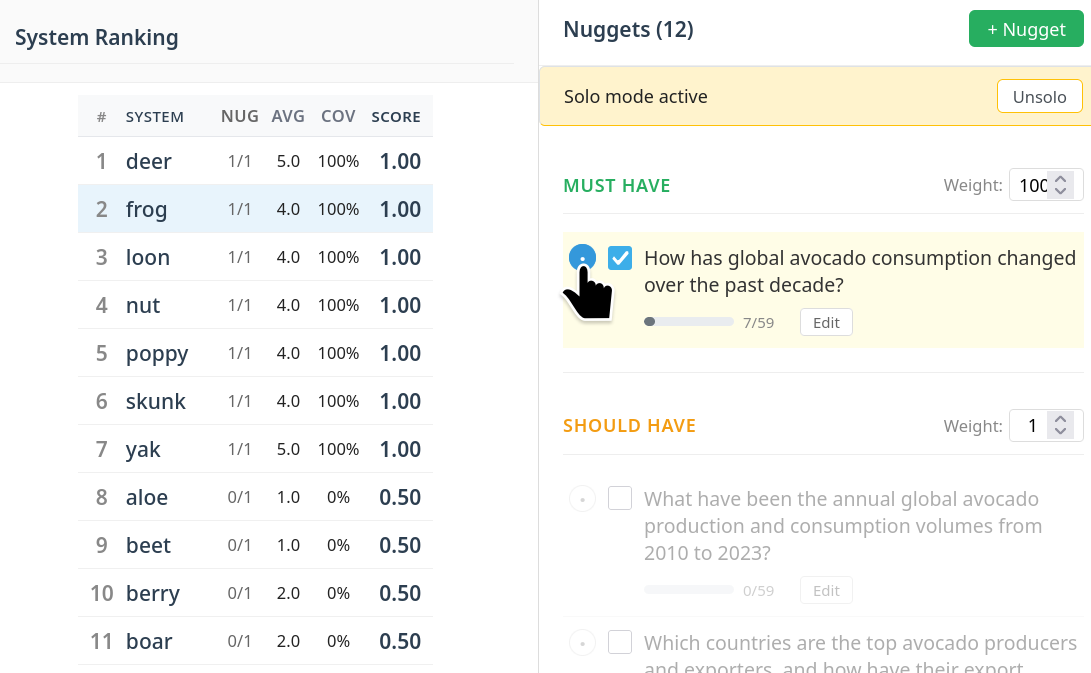}
\caption{Inspecting the nugget coverage and impact on the system ranking. Different functions allow experts to assess the impact that the collection of nuggets has together and in isolation. This helps the human expert to confirm that the result agrees with their informal impression of the system quality.}
\end{figure}

The goal of this stage is to ensure that the nugget bank reveals meaningful quality differences among the strongest systems. The expert should also confirm that low-ranked systems genuinely omit useful information, rather than being penalized because the nugget bank overlooked content they do contain.

Across these steps, the common pattern is that AI support remains downstream of a human-authored artifact, preserving human accountability while improving consistency and feedback.

\section{Avoiding Evaluation Tropes}\label{avoiding-evaluation-traps}

The prototype addresses several recurring failure modes in LLM evaluation by assigning each risky decision point to either the human expert, the LLM, or an explicit inspection step.

\paragraph{Avoiding Anchoring (Rubber-Stamp
Effect)}\label{avoiding-anchoring-rubber-stamp-effect}

The human identifies information before any LLM suggestion. Canonicalization only formalizes a human-authored artifact, so the human is not asked to accept or reject an AI proposal before forming their own judgment.

\paragraph{Avoiding Black-Box
Labeling}\label{avoiding-black-box-labeling}

Rather than asking humans for a single holistic quality score, the workflow asks them to create concrete, checkable artifacts. This makes the evaluation criterion visible as a set of nuggets instead of leaving it implicit inside an opaque label.

\paragraph{Avoiding Criteria Drift}\label{avoiding-criteria-drift}

The impact feedback preview reveals immediately when a nugget produces
unexpected grades. The human discovers interpretation issues during
creation, not after annotating an entire test set. The QC phase provides
aggregate diagnostics (universal and hard nugget counts) that flag
systematic problems.

\paragraph{Enabling Human Accountability}\label{enabling-human-accountability}

The LLM cannot unilaterally introduce evaluation criteria: every nugget exists because a human selected, wrote, categorized, or edited it. This makes accountability inspectable at the level of individual artifacts rather than only at the level of aggregate scores.

\paragraph{Maintaining Genuine Oversight}\label{maintaining-genuine-oversight}

Because grades are tied to individual nuggets and supporting matches, they can be inspected, audited, and revisited when systems or standards drift. The exported nugget bank therefore serves as a human-grounded reference that can be reused for method development, LLM evaluation, and deployment observability.

\section{Related Work}\label{related-work}

% \textbf{Nugget-based Evaluation.} 

Nugget-based evaluation has a two-decade history in information
retrieval, originating at TREC \cite{voorhees2003trec}. 

\citet{voorhees2003trec,nenkova2004evaluating} introduced nugget
pyramids for question answering evaluation. \citet{lin2006will} explored automated matching. \citet{pavlu2012nugget} use nuggets to evaluate information retrieval systems. \citet{sander2021exam} use multiple-choice questions with a Q/A system to evaluate long-form summaries.

Early nugget evaluations required manual matching by NIST assessors, making them expensive and difficult to scale. This work-intensive matching step hindered adoption despite the conceptual advantages of nugget-based assessment. Modern NLP and LLM capabilities make it feasible to automate the matching step while preserving human control over what to match \cite{sander2021exam,dietz2024workbench,farzi2024pencils,pradeep2024autonuggetizer,walden2025auto}.

\textbf{LLM-as-a-Judge.} \cite{faggioli2023perspectives,zheng2023judging} formalize the pattern of using LLMs for evaluation, but already discuss a range of issues. Subsequent work documented biases including position bias \cite{shi2025judging}, leniency bias \cite{pradeep2025great,farzi2024exampp}, and self-preferential bias \cite{liu2024narcissistic}.

\textbf{Human-AI Collaboration.} \citet{agudo2024impact} study
anchoring effects in AI-assisted labeling. \citet{shankar2024validates}
documented criteria drift in annotation tasks.

\textbf{Judge failure modes.} \citet{dietz2025principles} cataloged 14 failure
modes in LLM evaluation, including the Rubber-Stamp Effect and Black-Box
Labeling addressed here.

% \subsection{Limitations and Future
% Work}\label{limitations-and-future-work}

% \textbf{Current Limitations.} The prototype requires human expertise to
% identify relevant nuggets. Browser-based storage limits dataset size.
% There is no collaborative annotation support (single user per session).

% \textbf{Planned Extensions.} Pre-alignment of suggested nuggets across
% responses Farzi and colleagues 2026). Collaborative annotation with
% conflict resolution. Integration with continuous evaluation pipelines.

\section{Conclusion}\label{conclusion}

The question is not whether to include humans in LLM evaluation, but how
to include them effectively. The standard approaches, verify-and-correct
or independent labeling, either anchor humans to machine judgments or
leave them unsupported in high-variance tasks.

This prototype demonstrates a different division of labor. Humans
contribute where they excel (identifying what matters) while LLMs handle
what they do well (high-volume linguistic matching). The three-phase
workflow supports creation, calibration, and analysis. Exported nugget
banks enable reproducible, auditable evaluation.

The fix for circular LLM-as-a-Judge evaluation is not a better prompt. It is a
better division of labor. Three design principles readily translate to other domains where humans and AI jointly create evaluation criteria:
% \subsubsection{Key Design Decisions}\label{key-design-decisions}

\textbf{Human Initiative, LLM Assistance.} Put the human action before the machine suggestion. In other domains, this means asking experts to articulate their intent, evidence, constraint, or concern before using AI to clean up wording, normalize format, or apply the criterion at scale.

\textbf{Impact Feedback Before Commitment.} Show experts the downstream effect of their decisions while they can still revise them. Natural-language criteria may seem clear in isolation, but their consequences only become visible when applied to real cases; previewing those consequences helps experts refine criteria before they become part of an evaluation pipeline.

\textbf{Narrow, Verifiable LLM Tasks.} Assign the LLM tasks that can be checked directly, such as rewriting a human-authored artifact into a canonical form or matching that artifact against candidate outputs. Avoid asking the LLM to make open-ended normative decisions that should remain human responsibilities.

\bigskip

These principles suggest a broader alternative to fully automated evaluation: reserve intellectually demanding choices for accountable human experts, and use AI to make those choices easier to express, apply, inspect, and repeat.

% \subsection{References}\label{references}

% Agudo, U., and colleagues (2024). Anchoring effects in AI-assisted
% annotation. CHI 2024.

% Dietz, L., and colleagues (2025). Principles of LLM Judge Design. SIGIR
% Resource Track.

% Farzi, N., and colleagues (2026). Cross-response nugget alignment for
% efficient annotation. (In preparation)

% Lin, J., and Demner-Fushman, D. (2006). Will pyramids built of nuggets
% topple over? NAACL-HLT.

% Shankar, S., and colleagues (2024). Who validates the validators?
% Aligning LLM-assisted evaluation of LLM outputs with human preferences.
% ArXiv.

% Tversky, A., and Kahneman, D. (1974). Judgment under uncertainty:
% Heuristics and biases. Science 185(4157):1124-1131.

% Voorhees, E. M. (2003). Overview of the TREC 2003 question answering
% track. TREC.

% Zheng, L., and colleagues (2023). Judging LLM-as-a-judge with MT-Bench
% and Chatbot Arena. NeurIPS.

% \begin{center}\rule{0.5\linewidth}{0.5pt}\end{center}

\bibliographystyle{ACM-Reference-Format}
\bibliography{vulgen,bibio}

@inproceedings{lin2006will,
  title={Will pyramids built of nuggets topple over?},
  author={Lin, Jimmy and Demner-Fushman, Dina},
  booktitle={Proceedings of the Human Language Technology Conference of the NAACL, Main Conference},
  pages={383--390},
  year={2006}
}

@inproceedings{farzi2024exampp,
  title={Exam++: Llm-based answerability metrics for ir evaluation},
  author={Farzi, Naghmeh and Dietz, Laura},
  booktitle={Proceedings of LLM4Eval: The First Workshop on Large Language Models for Evaluation in Information Retrieval},
  year={2024}
}

@article{Zheng2023judging, title={Judging LLM-as-a-Judge with MT-Bench and Chatbot Arena}, url={http://arxiv.org/abs/2306.05685}, DOI={10.48550/arXiv.2306.05685}, abstractNote={Evaluating large language model (LLM) based chat assistants is challenging due to their broad capabilities and the inadequacy of existing benchmarks in measuring human preferences. To address this, we explore using strong LLMs as judges to evaluate these models on more open-ended questions. We examine the usage and limitations of LLM-as-a-judge, including position, verbosity, and self-enhancement biases, as well as limited reasoning ability, and propose solutions to mitigate some of them. We then verify the agreement between LLM judges and human preferences by introducing two benchmarks: MT-bench, a multi-turn question set; and Chatbot Arena, a crowdsourced battle platform. Our results reveal that strong LLM judges like GPT-4 can match both controlled and crowdsourced human preferences well, achieving over 80% agreement, the same level of agreement between humans. Hence, LLM-as-a-judge is a scalable and explainable way to approximate human preferences, which are otherwise very expensive to obtain. Additionally, we show our benchmark and traditional benchmarks complement each other by evaluating several variants of LLaMA and Vicuna. The MT-bench questions, 3K expert votes, and 30K conversations with human preferences are publicly available at https://github.com/lm-sys/FastChat/tree/main/fastchat/llm_judge.}, note={arXiv:2306.05685 [cs]}, number={arXiv:2306.05685}, publisher={arXiv}, author={Zheng, Lianmin and Chiang, Wei-Lin and Sheng, Ying and Zhuang, Siyuan and Wu, Zhanghao and Zhuang, Yonghao and Lin, Zi and Li, Zhuohan and Li, Dacheng and Xing, Eric P. and Zhang, Hao and Gonzalez, Joseph E. and Stoica, Ion}, year={2023}, month=dec }

@inproceedings{sander2021exam,
  title={EXAM: How to Evaluate Retrieve-and-Generate Systems for Users Who Do Not (Yet) Know What They Want.},
  author={Sander, David P and Dietz, Laura},
  booktitle={DESIRES},
  pages={136--146},
  year={2021}
}

@article{fok2024search,
  author    = {Fok, Raymond and Weld, Daniel S},
  journal   = {AI Magazine},
  number    = {3},
  pages     = {317--332},
  publisher = {Wiley Online Library},
  title     = {In search of verifiability: Explanations rarely enable complementary performance in AI-advised decision making},
  volume    = {45},
  year      = {2024}
}

@inproceedings{clarke2024llm,
  author    = {Clarke, Charles L. A.  and Dietz, Laura},
  booktitle = {EVIA 2025: Proceedings of the Tenth International Workshop on Evaluating Information Access (EVIA 2025), a Satellite Workshop of the NTCIR-18 Conference, June 10-13, 2025, Tokyo, Japan},
  doi       = {10.20736/0002002105},
  pages     = {1--5},
  title     = {LLM-based relevance assessment still can't replace human relevance assessment},
  year      = {2025}
}

@inproceedings{liu2024narcissistic,
  author    = {Yiqi Liu and Nafise Sadat Moosavi and Chenghua Lin},
  title     = {LLMs as Narcissistic Evaluators: When Ego Inflates Evaluation Scores},
  booktitle = {Findings of the Association for Computational Linguistics (ACL) 2024},
  year      = {2024},
  note      = {Investigates bias in LLM-based evaluation metrics favoring their own outputs},
  url       = {https://aclanthology.org/2024.findings-acl.753/}
}

@unpublished{pradeep2024autonuggetizer,
  author    = {Ronak Pradeep and Nandan Thakur and Shivani Upadhyay and Daniel Campos and Nick Craswell and Jimmy Lin},
  title     = {Initial Nugget Evaluation Results for the {TREC} 2024 RAG Track with the AutoNuggetizer Framework},
  year      = {2024},
  note      = {ArXiv preprint},
  url       = {https://arxiv.org/abs/2411.09607}
}

@inproceedings{dietz2024workbench,
  title={A workbench for autograding retrieve/generate systems},
  author={Dietz, Laura},
  booktitle={Proceedings of the 47th International ACM SIGIR Conference on Research and Development in Information Retrieval},
  pages={1963--1972},
  year={2024}
}

@inproceedings{voorhees2003trec,
  author    = {Ellen M. Voorhees},
  title     = {Overview of the {TREC} 2003 Question Answering Track},
  booktitle = {Proceedings of the Twelfth Text REtrieval Conference (TREC 2003)},
  year      = {2003},
  address   = {Gaithersburg, Maryland},
  publisher = {NIST}
}

@inproceedings{pavlu2012nugget,
  author    = {Virgil Pavlu and Shahzad Rajput and Peter B. Golbus and Javed A. Aslam},
  title     = {{IR} System Evaluation Using Nugget-Based Test Collections},
  booktitle = {Proceedings of the Fifth ACM International Conference on Web Search and Data Mining (WSDM 2012)},
  year      = {2012},
  pages     = {393--402},
  address   = {Seattle, Washington},
  publisher = {ACM}
}

@inproceedings{shi2025judging,
  title={Judging the judges: A systematic study of position bias in llm-as-a-judge},
  author={Shi, Lin and Ma, Chiyu and Liang, Wenhua and Diao, Xingjian and Ma, Weicheng and Vosoughi, Soroush},
  booktitle={Proceedings of the 14th International Joint Conference on Natural Language Processing and the 4th Conference of the Asia-Pacific Chapter of the Association for Computational Linguistics},
  pages={292--314},
  year={2025}
}

@inproceedings{li2026dogmatiq,
  author    = {Li, Bryan and Walden, William and Hou, Yu and Liu, Gabrielle Kaili-May and Lawrie, Dawn and Mayfield, James and Yang, Eugene and Callison-Burch, Chris and Dietz, Laura},
  title     = {DoGMaTiQ: Automated Generation of Question-and-Answer Nuggets for Report Evaluation},
  booktitle = {Proceedings of the 2026 ACM SIGIR International Conference on the Theory of Information Retrieval (ICTIR '26)},
  year      = {2026},
  month     = jul,
  address   = {Melbourne, VIC, Australia},
  publisher = {ACM}
}

@inproceedings{dietz2026too,
  author    = {Dietz, Laura and Farzi, Naghmeh and Yang, Eugene and Lawrie, Dawn},
  title     = {Too Many Questions: Deriving Concise and Effective Nugget Banks},
  booktitle = {Proceedings of the 49th International ACM SIGIR Conference on Research and Development in Information Retrieval (SIGIR '26)},
  year      = {2026},
  month     = jul,
  date      = {July 20--24, 2026},
  address   = {Melbourne, VIC, Australia},
  publisher = {ACM}
}

@inproceedings{shankar2024validates,
  title={Who validates the validators? aligning llm-assisted evaluation of llm outputs with human preferences},
  author={Shankar, Shreya and Zamfirescu-Pereira, JD and Hartmann, Bj{\"o}rn and Parameswaran, Aditya and Arawjo, Ian},
  booktitle={Proceedings of the 37th Annual ACM Symposium on User Interface Software and Technology},
  pages={1--14},
  year={2024}
}

@article{agudo2024impact,
  title={The impact of AI errors in a human-in-the-loop process},
  author={Agudo, Uju{\'e} and Liberal, Karlos G and Arrese, Miren and Matute, Helena},
  journal={Cognitive Research: Principles and Implications},
  volume={9},
  number={1},
  pages={1},
  year={2024},
  publisher={Springer}
}

@article{tversky1974judgment,
  title={Judgment under Uncertainty: Heuristics and Biases: Biases in judgments reveal some heuristics of thinking under uncertainty.},
  author={Tversky, Amos and Kahneman, Daniel},
  journal={science},
  volume={185},
  number={4157},
  pages={1124--1131},
  year={1974},
  publisher={American association for the advancement of science}
}

@inproceedings{walden2025auto,
  title={Auto-argue: Llm-based report generation evaluation},
  author={Walden, William and Mason, Marc and Weller, Orion and Dietz, Laura and Conroy, John and Molino, Neil and Recknor, Hannah and Li, Bryan and Liu, Gabrielle Kaili-May and Hou, Yu and others},
  booktitle={SIGIR},
  year={2026}
}

@inproceedings{nenkova2004evaluating,
  title={Evaluating content selection in summarization: The pyramid method},
  author={Nenkova, Ani and Passonneau, Rebecca J},
  booktitle={Proceedings of the human language technology conference of the north american chapter of the association for computational linguistics: Hlt-naacl 2004},
  pages={145--152},
  year={2004}
}

@inproceedings{dietz2025principles,
  author    = {Dietz, Laura and Zendel, Oleg and Bailey, Peter and Clarke, Charles L. A. and Cotterill, Ellese and Dalton, Jeff and Hasibi, Faegheh and Sanderson, Mark and Craswell, Nick},
  title     = {Principles and Guidelines for the Use of {LLM} Judges},
  booktitle = {Proceedings of the 2025 International ACM SIGIR Conference on Innovative Concepts and Theories in Information Retrieval (ICTIR '25)},
  year      = {2025},
  doi       = {10.1145/3731120.3744588}
}

@inproceedings{dietz2025tropes,
  title={Principles and Guidelines for the Use of LLM Judges},
  author={Dietz, Laura and Zendel, Oleg and Bailey, Peter and Clarke, Charles and Cotterill, Ellese and Dalton, Jeff and Hasibi, Faegheh and Sanderson, Mark and Craswell, Nick},
  booktitle={Proceedings of the 11th ACM SIGIR / The 15th International Conference on Innovative Concepts and Theories in Information Retrieval},
  year={2025}
}

@inproceedings{dietz2026insider,
  author    = {Dietz, Laura and Li, Bryan and Yang, Eugene and Lawrie, Dawn and Walden, William and Mayfield, James},
  title     = {Insider Knowledge: How Much Can {RAG} Systems Gain from Evaluation Secrets?},
  booktitle = {Proceedings of the 48th European Conference on Information Retrieval (ECIR 2026)},
  year      = {2026},
  note      = {arXiv:2601.13227}
}

@inproceedings{farzi2024pencils,
  author    = {Farzi, Naghmeh and Dietz, Laura},
  title     = {Pencils Down! Automatic Rubric-based Evaluation of Retrieve/Generate Systems},
  booktitle = {Proceedings of the 2024 ACM SIGIR International Conference on the Theory of Information Retrieval (ICTIR '24)},
  year      = {2024},
  doi       = {10.1145/3664190.3672511}
}

@inproceedings{faggioli2023perspectives,
  author    = {Faggioli, Guglielmo and Dietz, Laura and Clarke, Charles L. A. and Demartini, Gianluca and Hagen, Matthias and Hauff, Claudia and Kando, Noriko and Kanoulas, Evangelos and Potthast, Martin and Stein, Benno and others},
  title     = {Perspectives on Large Language Models for Relevance Judgment},
  booktitle = {Proceedings of the 2023 ACM SIGIR International Conference on Theory of Information Retrieval},
  year      = {2023},
  pages     = {39--50}
}

@inproceedings{pradeep2025great,
  title     = {The Great Nugget Recall: Automating Fact Extraction and {RAG} Evaluation with Large Language Models},
  author    = {Pradeep, Ronak and Thakur, Nandan and Upadhyay, Shivani and Campos, Daniel and Craswell, Nick and Soboroff, Ian and Dang, Hoa Trang and Lin, Jimmy},
  booktitle = {Proceedings of the 48th International ACM SIGIR Conference on Research and Development in Information Retrieval},
  pages     = {180--190},
  year      = {2025}
}

\FloatBarrier
\eject{}
\appendix
\section{Appendix: Walkthrough}\label{appendix-walkthrough}

This appendix demonstrates the prototype through three scenarios that
illustrate the key design principles.

\subsection{Human Goes First, AI
Supports}\label{a.1-human-goes-first-ai-supports}

This walkthrough shows how the interface ensures the human forms their
judgment before any LLM involvement.

\begin{figure}
\centering
\includegraphics[width=\columnwidth,keepaspectratio,alt={Step 1-1a}]{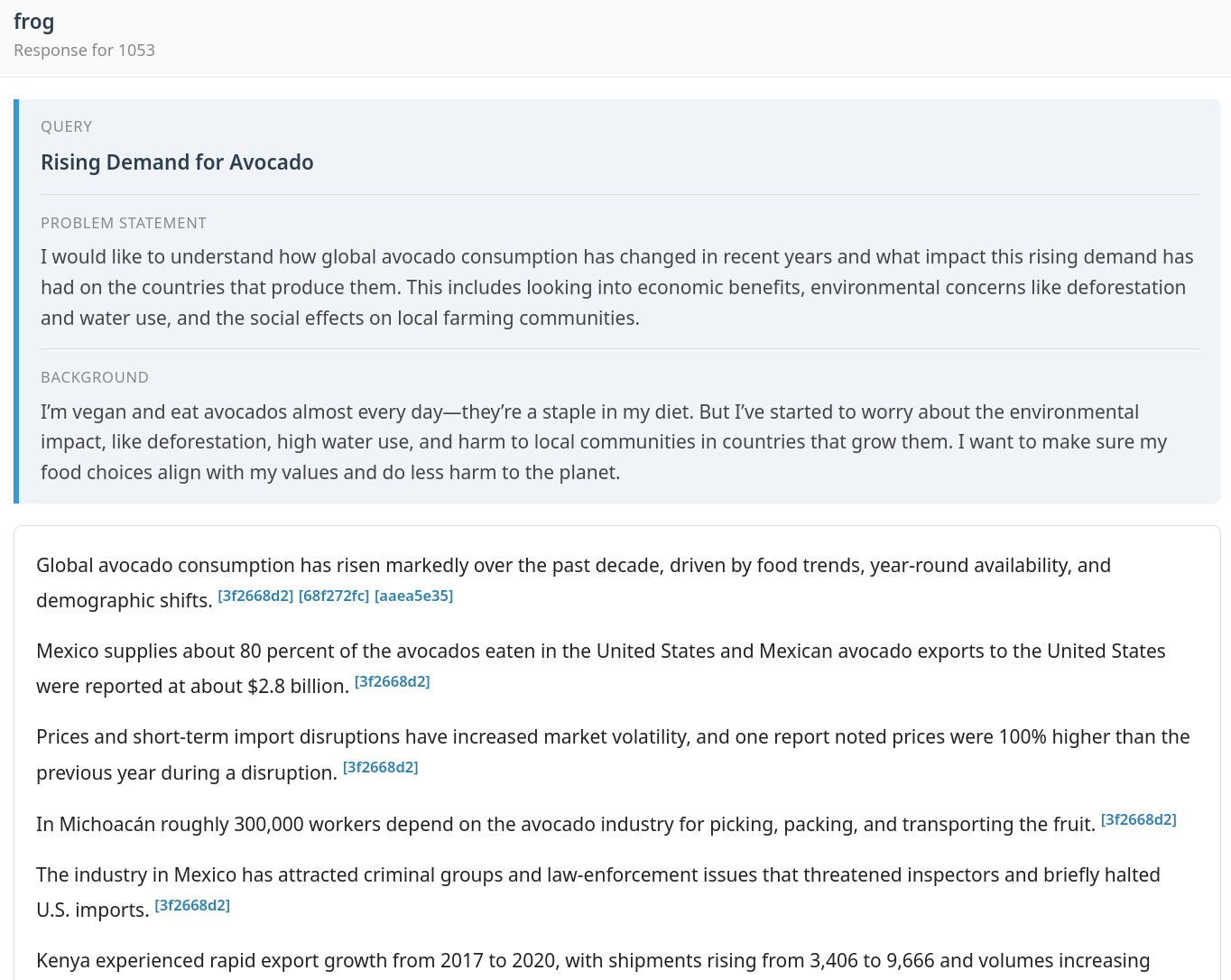}
\caption{\textbf{Step 1: Human reads the query and report.} The annotator sees the query ``Rising Demand for Avocado'' with its problem statement and background context. Below, the system response describes global avocado trends, Mexico's market share, and various impacts. The human reads and forms their own understanding of what information matters. No LLM has been invoked.}
\end{figure}

\begin{figure}
\centering
\includegraphics[width=1\columnwidth,keepaspectratio,alt={Step 1-1}]{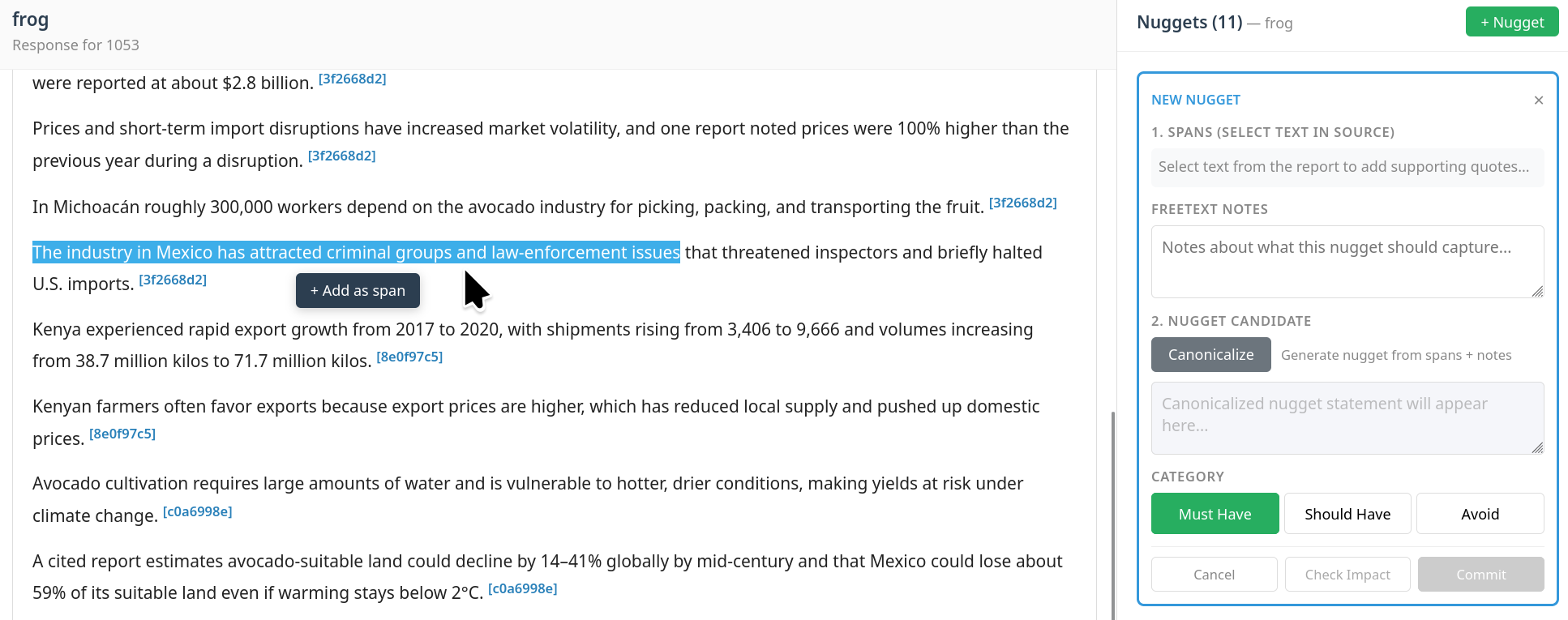}
\caption{\textbf{Step 2: Human selects text.} The annotator highlights text: ``The industry in Mexico has attracted criminal groups and law-enforcement issues.'' A popup appears offering ``+ Add as span.'' The draft card opens on the right. The LLM has still not been involved. The human has already identified what they consider important.}
\end{figure}

\begin{figure}
\centering
\includegraphics[width=1\columnwidth,keepaspectratio,alt={Step 1-2}]{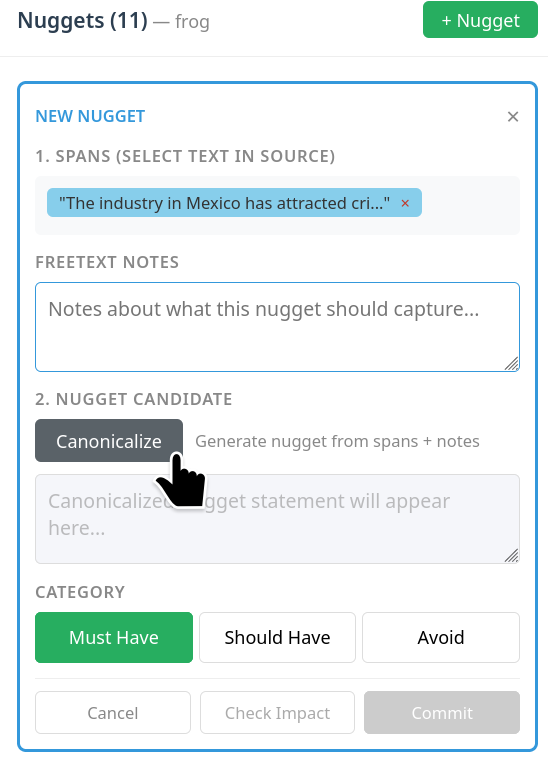}
\caption{\textbf{Step 3: Span added, ready for notes.} The draft card shows the selected span as a chip. The Canonicalize button is visible but not yet clicked. At this point, the human's conceptual judgment is already formed.}
\end{figure}

\begin{figure}
\centering
\includegraphics[width=1\columnwidth,keepaspectratio,alt={Step 1-3}]{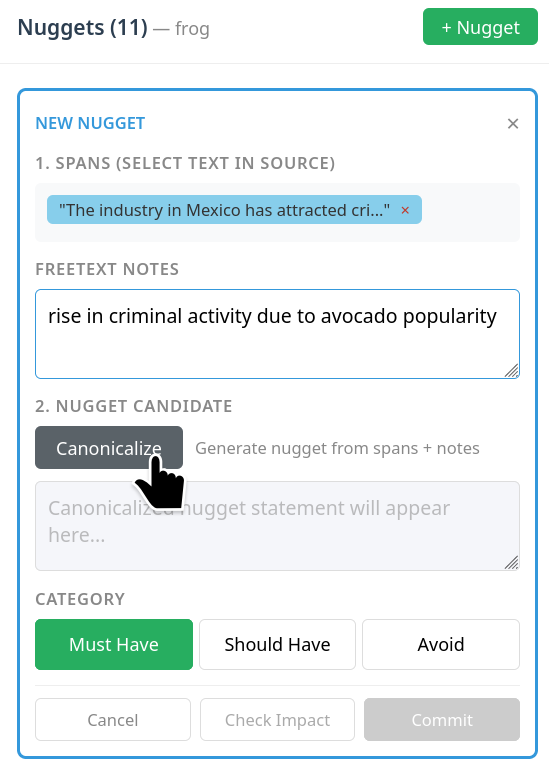}
\caption{\textbf{Step 4: Human adds context.} The annotator types free-text notes: ``rise in criminal activity due to avocado popularity.'' These notes describe their intent before any machine involvement.}
\end{figure}

\begin{figure}
\centering
\includegraphics[width=1\columnwidth,keepaspectratio,alt={Step 1-4}]{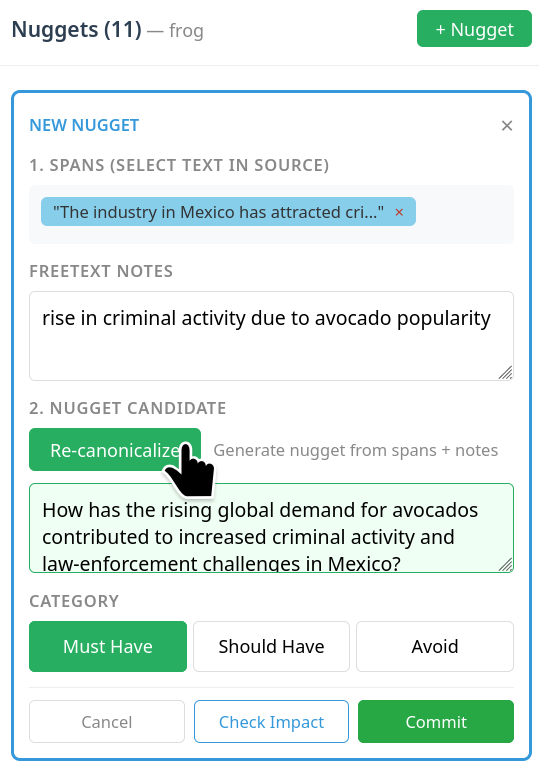}
\caption{\textbf{Step 5: LLM formalizes.} After clicking Canonicalize, the LLM generates: ``How has the rising global demand for avocados contributed to increased criminal activity and law-enforcement challenges in Mexico?'' The LLM's role is formalization, not proposal. The human decided this information matters; the LLM helped express it as a verifiable question.}
\end{figure}

\begin{figure}
\centering
\includegraphics[width=1\columnwidth,keepaspectratio,alt={Step 1-5}]{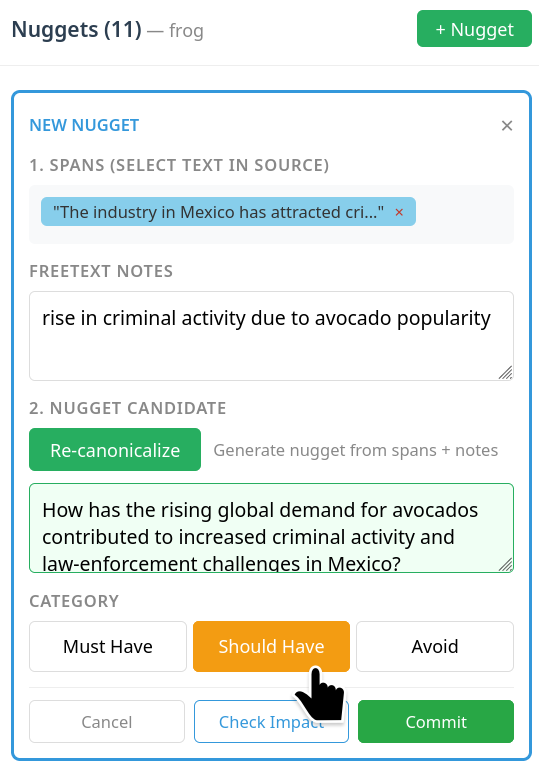}
\caption{\textbf{Step 6: Human chooses category.} The annotator selects ``Should Have'' as the category. The importance judgment is entirely human.}
\end{figure}

\FloatBarrier

\subsection{Feedback Loop via Check
Impact}\label{a.2-feedback-loop-via-check-impact}

This walkthrough shows how real-time feedback helps the annotator refine
nuggets before committing.

\begin{figure}
\centering
\includegraphics[width=1\columnwidth,keepaspectratio,alt={Step 2-2}]{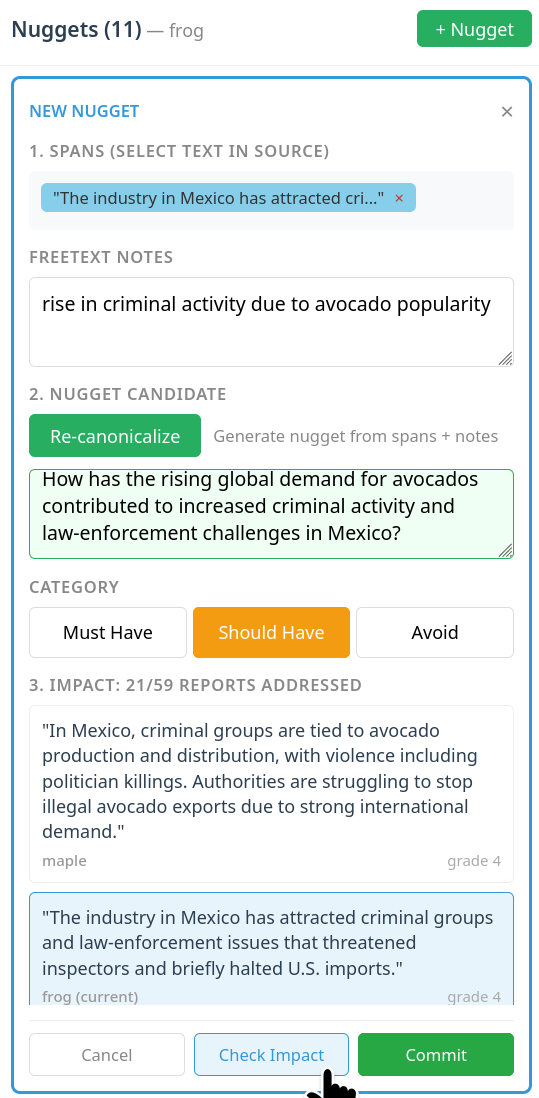}
\caption{\textbf{Step 1: Impact results with quotes.} After clicking Check Impact, the interface shows ``21/59 REPORTS ADDRESSED'' with supporting quotes. System ``maple'' (grade 4) shows: ``In Mexico, criminal groups are tied to avocado production and distribution, with violence including politician killings\ldots{}'' The annotator sees exactly how this nugget grades and why.}
\end{figure}

\begin{figure}
\centering
\includegraphics[width=1\columnwidth,keepaspectratio,alt={Step 2-4}]{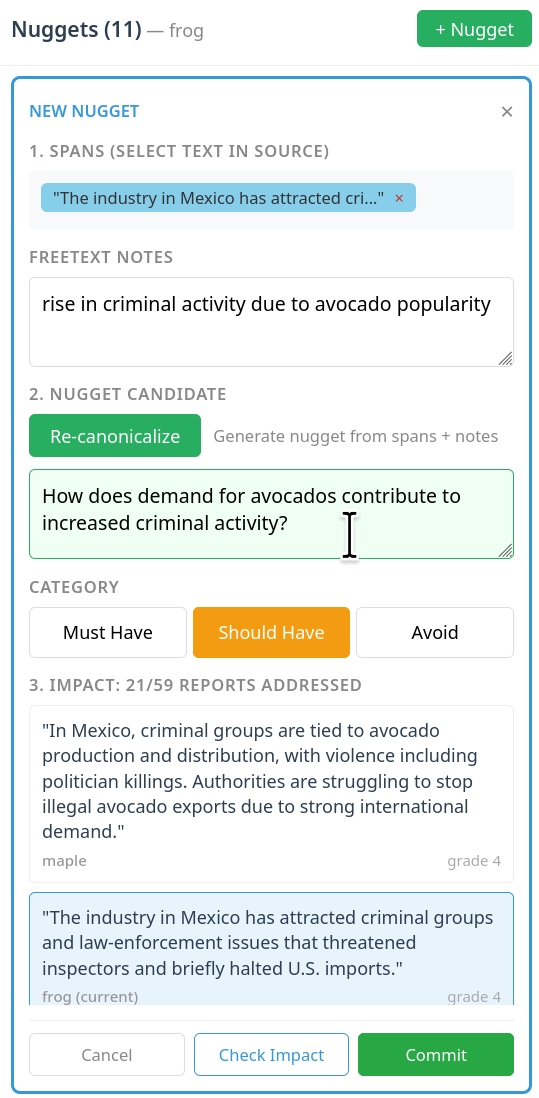}
\caption{\textbf{Step 2: Refining based on feedback.} Based on the preview, the annotator edits the nugget text to be more concise: ``How does demand for avocados contribute to increased criminal activity?'' This demonstrates the feedback loop: the annotator saw how the nugget performed and refined it.}
\end{figure}

\begin{figure}
\centering
\includegraphics[width=1\columnwidth,keepaspectratio,alt={Step 2-5}]{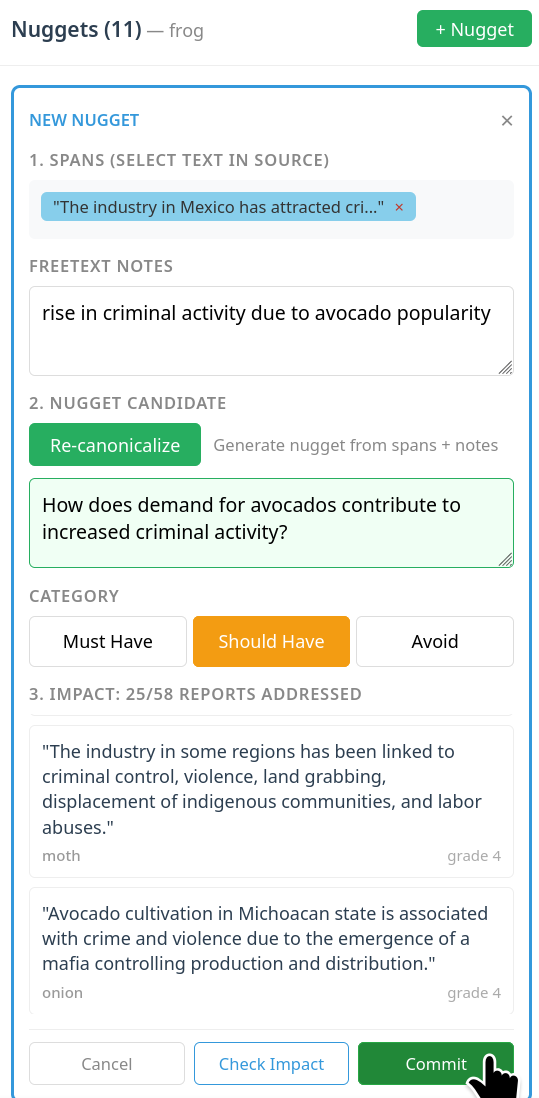}
\caption{\textbf{Step 3: Final verification.} The updated nugget shows ``25/58 REPORTS ADDRESSED'' with quotes from additional systems. The annotator clicks Commit, satisfied that the nugget discriminates effectively.}
\end{figure}

\FloatBarrier

\subsection{QC Phase for Improving Nugget Bank
Quality}\label{a.3-qc-phase-for-improving-nugget-bank-quality}

This walkthrough shows how the annotator tunes and validates the nugget
bank.

\begin{figure}
\centering
\includegraphics[width=1\columnwidth,keepaspectratio,alt={Step 3-1}]{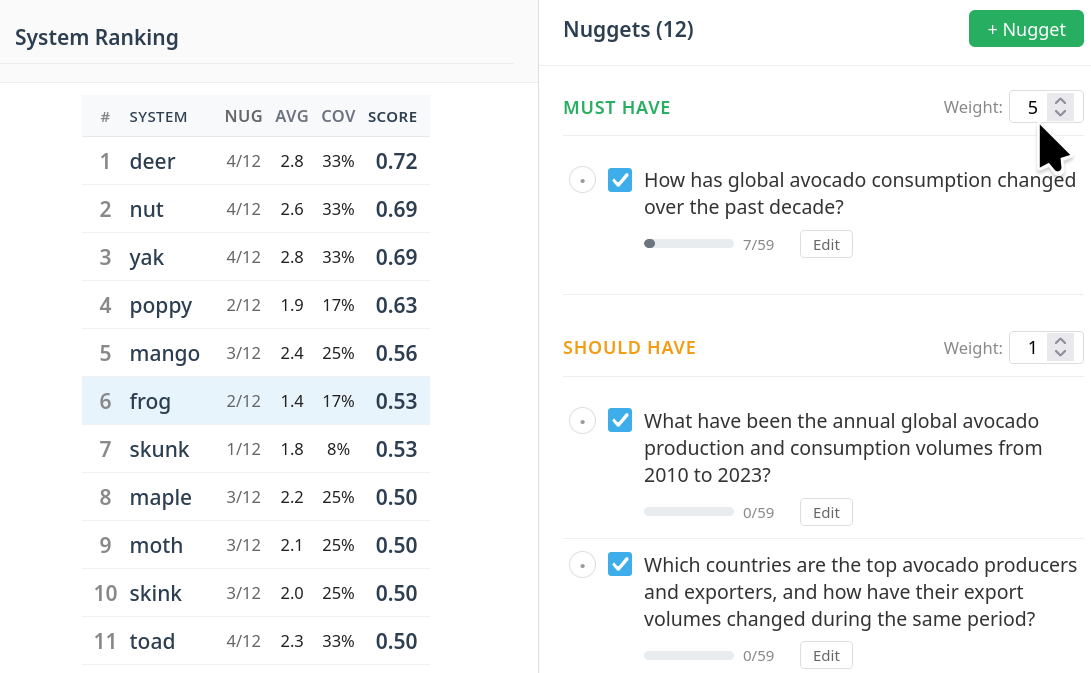}
\caption{\textbf{Step 1: QC phase with weight controls.} The QC phase shows the System Ranking table with columns for rank, system, nuggets satisfied (NUG), average grade (AVG), coverage (COV), and weighted score (SCORE). Category weights are adjustable (Must Have: 5, Should Have: 1).}
\end{figure}

\begin{figure}
\centering
\includegraphics[width=1\columnwidth,keepaspectratio,alt={Step 3-2}]{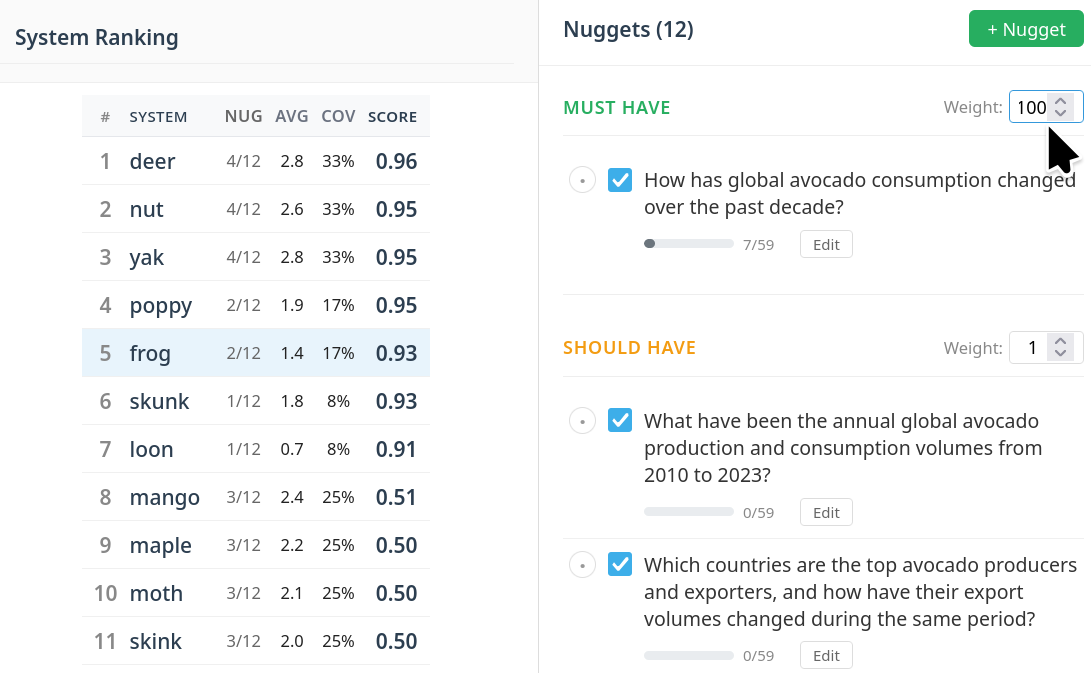}
\caption{\textbf{Step 2: Weight adjustment changes rankings.} Increasing Must Have weight to 100 causes immediate ranking changes. Scores compress and reorder. This reveals how sensitive the ranking is to category weights.}
\end{figure}

\begin{figure}
\centering
\includegraphics[width=1\columnwidth,keepaspectratio,alt={Step 3-3}]{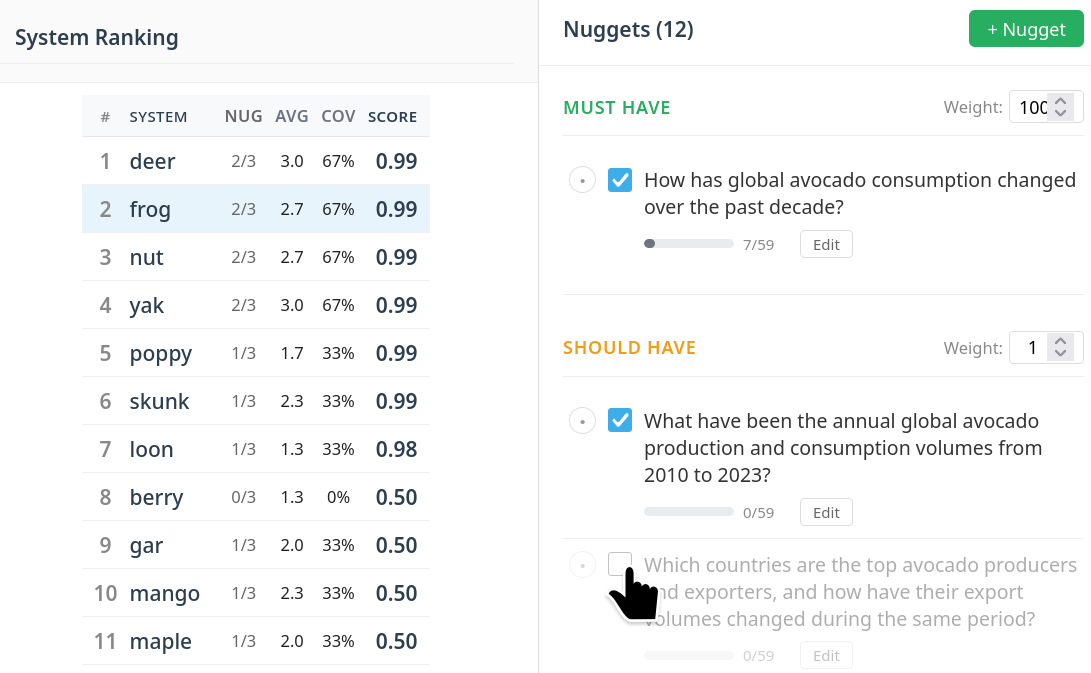}
\caption{\textbf{Step 3: Disable a nugget.} Unchecking a nugget removes it from scoring. The NUG column changes from ``4/12'' to ``2/3.'' This allows testing whether specific nuggets contribute meaningfully.}
\end{figure}

\begin{figure}
\centering
\includegraphics[width=1\columnwidth,keepaspectratio,alt={Step 3-4}]{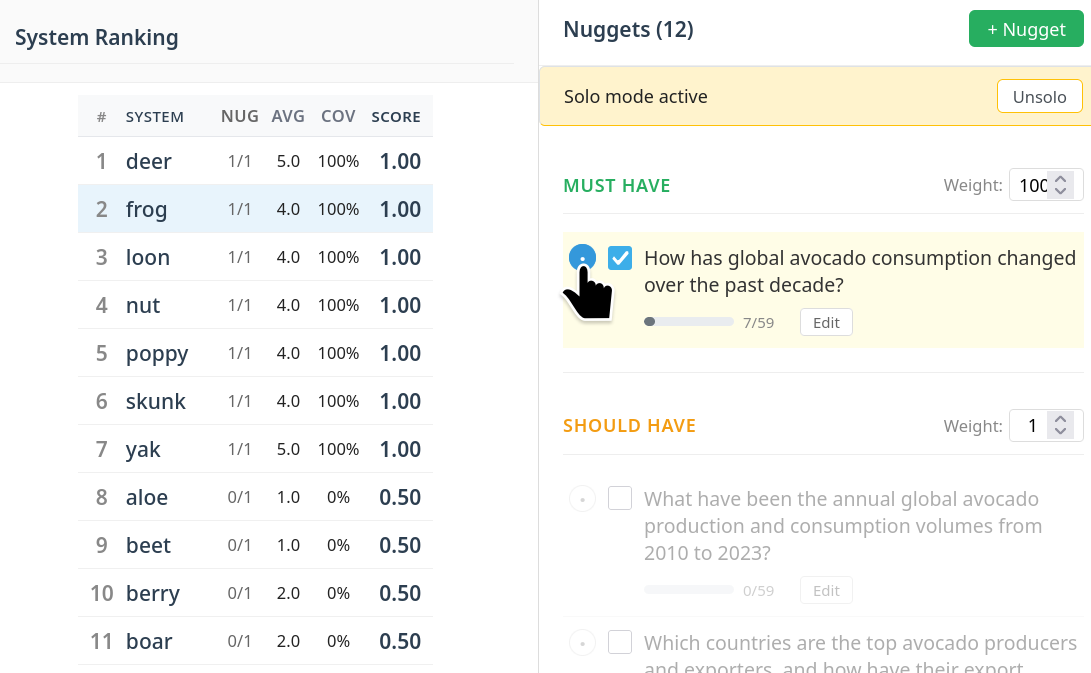}
\caption{\textbf{Step 4: Solo mode.} Solo mode isolates a single nugget. Systems split cleanly: those addressing the nugget score 1.00; others score 0.50. This reveals the nugget's discriminative power.}
\end{figure}

\begin{figure}
\centering
\includegraphics[width=1\columnwidth,keepaspectratio,alt={Step 3-5}]{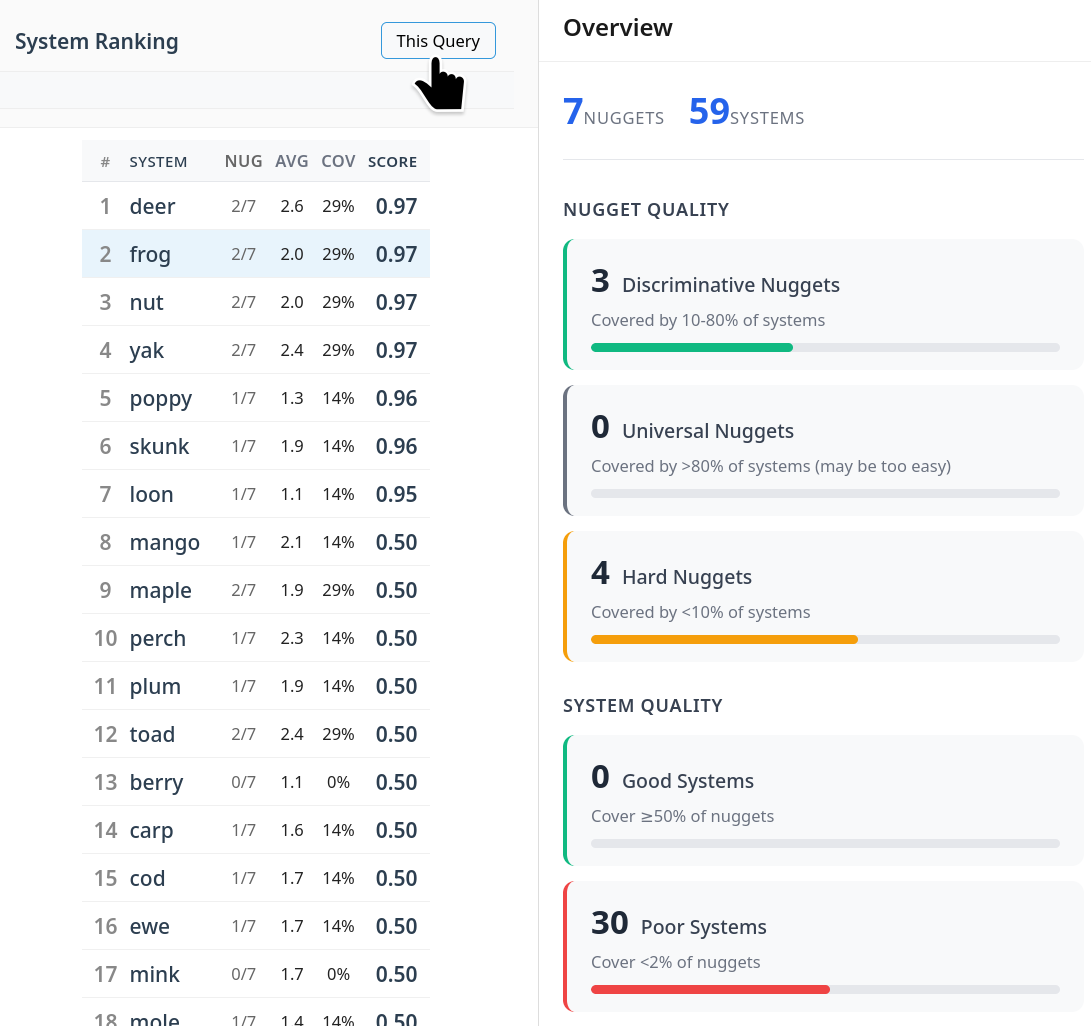}
\caption{\textbf{Step 5: Observe phase diagnostics.} The Observe phase shows aggregate statistics: 7 Nuggets, 59 Systems. Nugget Quality: 3 Discriminative (10--80\% coverage), 0 Universal, 4 Hard. System Quality: 0 Good Systems, 30 Poor Systems. The 4 Hard nuggets suggest the bank may be too strict.}
\end{figure}

\begin{figure}
\centering
\includegraphics[width=1\columnwidth,keepaspectratio,alt={Step 3-6}]{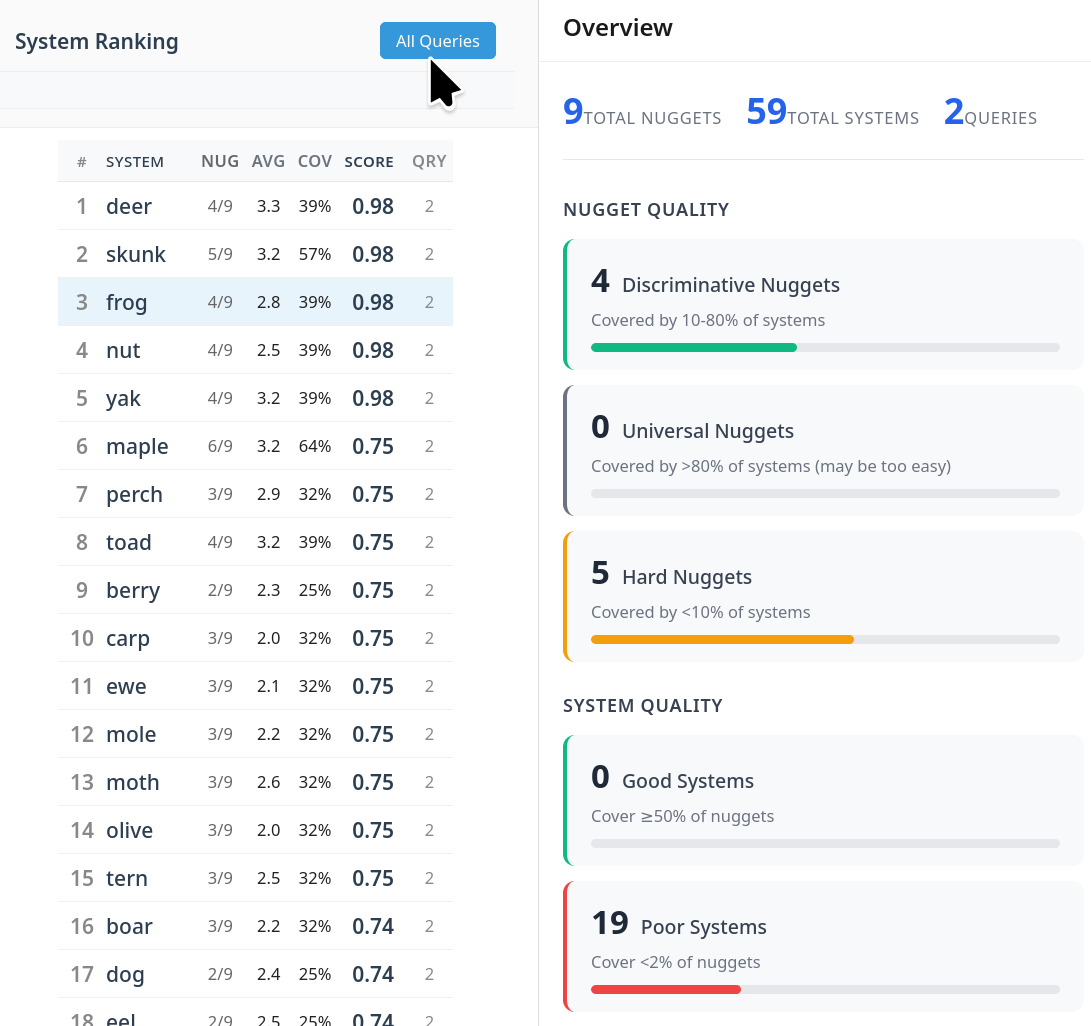}
\caption{\textbf{Step 6: Cross-query view.} ``All Queries'' shows macro-averaged statistics across 2 queries: 9 Total Nuggets, 4 Discriminative, 5 Hard. A QRY column shows per-system query coverage. This reveals which systems perform consistently.}
\end{figure}

\FloatBarrier

\subsection{Summary of Principles
Demonstrated}\label{a.4-summary-of-principles-demonstrated}
\begin{table}[ht]
\centering
\begin{tabular}{@{}
  >{\raggedright\arraybackslash}p{(\linewidth - 4\tabcolsep) * \real{0.3095}}
  >{\raggedright\arraybackslash}p{(\linewidth - 4\tabcolsep) * \real{0.2619}}
  >{\raggedright\arraybackslash}p{(\linewidth - 4\tabcolsep) * \real{0.4286}}@{}}
\toprule
Walkthrough & Principle & How Demonstrated \\
\midrule
A.1 & Human initiative & Selection and notes before canonicalization \\
A.1 & LLM assists, does not propose & Canonicalize formalizes human-identified information \\
A.1 & No anchoring & Human judgment formed before seeing LLM output \\
A.2 & Feedback before commit & Check Impact reveals grades and quotes instantly \\
A.2 & Avoiding criteria drift & Refinement happens during creation \\
A.2 & Verifiable matching & Quotes show exactly why grades were assigned \\
A.3 & Interactive calibration & Real-time weight adjustment \\
A.3 & Hypothesis testing & Disable and solo individual nuggets \\
A.3 & Aggregate diagnostics & Discriminative, universal, and hard nugget counts \\
\bottomrule
\end{tabular}
\caption{Walkthrough principles}
\end{table}

\end{document}